# Determination of Trace Moisture Content in Dimethylacetamide by Gas Chromatography

Yong-A Choe[†], Chan-Hyon Han[†], Kye-Ryong Sin[*,‡]

[†]Analytical Research Institute and [‡]Faculty of Chemistry, **Kim Il Sung** University, Pyongyang, Democratic People's Republic of Korea

**ABSTRACT:** Dimethylacetamide (DMA) is used as a solvent in polymer processing and moisture in DMA affects the morphology and mechanical properties of the polymer products. This paper describes a method for determination of trace moisture content in DMA by acetylene production - gas chromatography. In the condition of 30℃ and ultrasonic dispersion, moisture in DMA sample was reacted with calcium carbide and its product of acetylene was measured by gas chromatography. By this way, trace moisture in DMA was determined with RSD of 2.272% and limit of measurement was $2 \cdot 10^{-3}$%.

■ **INTRODUCTION**

The effect of moisture on the morphology and mechanical properties of different polymers has been extensively studied by a number of researchers[1-10]. In some cases, a specified amount of moisture content is required to achieve a certain property of polymer products[11-12]. Therefore, it is essential to check the moisture content of the compounding ingredients and solvents for polymer processing, along with the other quality control parameters.

By this time many methods to determine the trace moisture in organic solvents were presented [13-21], but those methods need complicated apparatus, materials and reagents and long analytical time. For instance, Karl Fischer (KF) titration[13], which is widely used for chemicals, is not suitable for some raw materials for polymer processing because they are insoluble in the KF solution. In this case, volatile loss on heating is measured[14,15], which gives the total amount of volatiles including the moisture, but not the exact moisture content. Another technique for moisture determination is the azeotropic reflux, but it can be applicable for samples with the moisture content more than 1%.

Dimethylacetamide (DMA) is widely used as solvent for polymer processing and also as reagent for metal catalysis, dehydration, reduction and other organic synthesis.[22,23]

The present article describes a method to determine the trace moisture in DMA by gas chromatography (GC) using the flame ionization detector (FID) with acetylene production.

■ **EXPERIMENTAL DETAILS**

As apparatus used were gas chromatography (GC-14B, Shimadzu), micro-injector (1 μL), analytical balance, micropipette (200~1000 μL), microwave reactor (KM-410L), screw-cap vial. All chemicals such as calcium carbide and dimethylacetamide were analytical reagent grade.

In acetylene production-moisture content measurement, trace water in DMA sample reacts with calcium carbide to produce acetylene. At that time, this acetylene is distributed in liquid-phase.

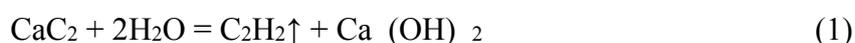
$$CaC_2 + 2H_2O = C_2H_2\uparrow + Ca(OH)_2 \qquad (1)$$

As long as water is present in sample, the acetylene was generated according to reaction (1). Thus, the trace amount of moisture can be determined by measuring the acetylene dissolved in the sample with gas chromatography.

At first, 600 μL of DMA was put into the screw-cap vial with a micropipette. And then, calcium carbide was put into the vial and was closed with a cap and then the vial was shaken. After 10 min, the supernatant liquid 1 μL was taken with a micro-injector and injected in the gas chromatography [Carrier gas (nitrogen) flow rate 0.61 ml/min, injection port temperature 200 ℃, initial temperature 80 ℃ (1 min), final temperature 180 ℃ (1 min), program rate 5 ℃/min, detector temperature 200 ℃, alkylphenoloxide (APO) column (3m×3mm), flame ionization detector (FID)] to measure acetylene.

■ **RESULTS AND DISCUSSION**

**Acetylene-production according to the measuring (ultrasonic dispersion) time.** Calcium carbide was put into the vial of DMA with certain amount of moisture and mixed by using ultrasonic reactor in 30 ℃. By the gas chromatography, the sample was measured according to the time of ultrasonic dispersion.

Change of the peak area of acetylene according to the time was showed in Figure 1. The time necessary to finish the reaction between calcium carbide and moisture in DMA to produce acetylene was about 200 min, while the peak area was changed. In this case, the calibration curve method was unable to apply. Thus, the method of standard addition was used.

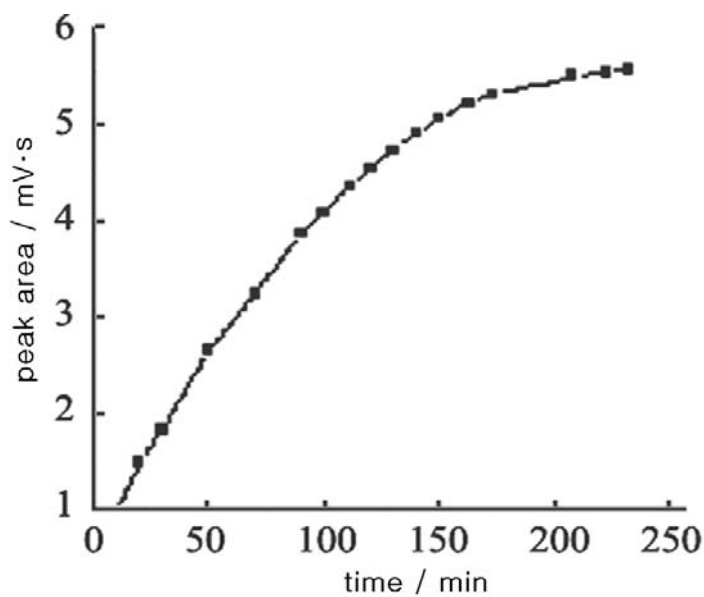

Figure 1. Change of peak area of acetylene according to the measuring time

**Determination of moisture content by the method of standard addition.** 600 μL of DMA was put into the each of two screw-cap vials, and the mass of them was measured, respectively. After 1 μL of water was added into one of two vials, the mass of it was determined. The time of measuring the one sample was 10 min, the suitable amount of calcium carbide was added to the two samples leaving an interval of 10 min. Next, the vials were closed and dispersed in microwave reactor. After it, leaving an interval of 10 min, 1 μL from each of two samples was injected into gas chromatography and measured the amount of produced acetylene. In this way, the reaction time for trace moisture in DMA and calcium carbide in each of two vials was kept equal.

By using the peak area of acetylene in the gas chromatogram, the amount of trace moisture in DMA was calculated as follows.

The mass of water in the sample (g), that is, $m_x$ was calculated according to Equation 2.

$$\frac{m_x}{m} : A_x = \frac{m_x + m_1}{m + m_1} : A_s \qquad (2)$$

where $m_x$: mass of water in the sample (g), $m$: mass of the sample (g), $m_1$: mass of water added to the sample (g), $A_x$: the peak area of acetylene in the sample (mV·s), $A_s$: peak area of acetylene in the water-added sample (mV·s).

And then, the moisture content ($x$) of the sample was determined according to Equation 3.

$$x(\%) = \frac{m_x}{m} \times 100 \qquad (3)$$

**Effects of the response temperature and ultrasonic dispersion to the acetylene-production.**
The measurement of peak area of acetylene in liquid phase was carried out in the different temperatures (Figure 2). Under 30 ℃ and over 40 ℃, the peak area of acetylene in the sample was small and was certain in 30 ~ 40 ℃. The effect of the response temperature to the acetylene-production was same in the primitive sample ( before adding water ) and in the water (1.0 μL) added –sample.

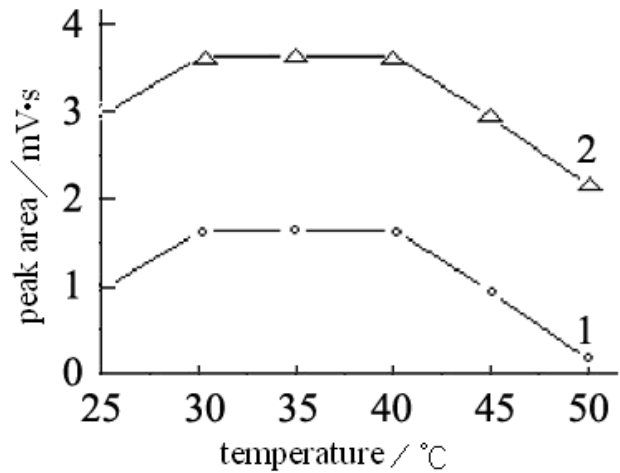

Figure 2. Effect of the response temperature to the acetylene-production
(1: primitive sample, 2: water added sample)

During the moisture measurement of all the samples, the temperature was fixed 30℃, considering the air temperature. Thus, the effect of the response temperature to the acetylene-production can be ignored.

The effect of ultrasonic dispersion to the acetylene-production was tested in 30℃ (Figure 3).

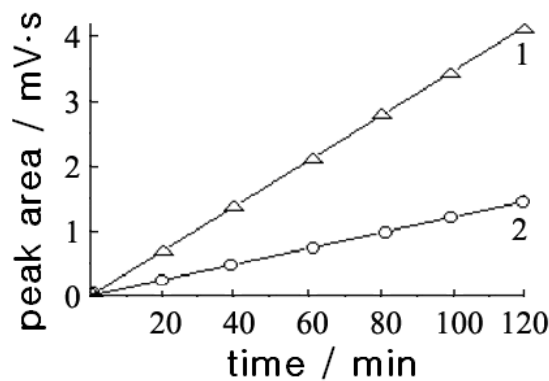

Figure 3. Effect of ultrasonic dispersion to the acetylene-production
(1: ultrasonic dispersed sample, 2: un-dispersed sample)

Figure 3 shows that ultrasonic dispersion can offer the short measuring time by making fast acetylene production.

The moisture content of the sample was not changed according to the measuring time by the method of standard addition under the conditions of 30℃ and ultrasonic dispersion (Figure 4).

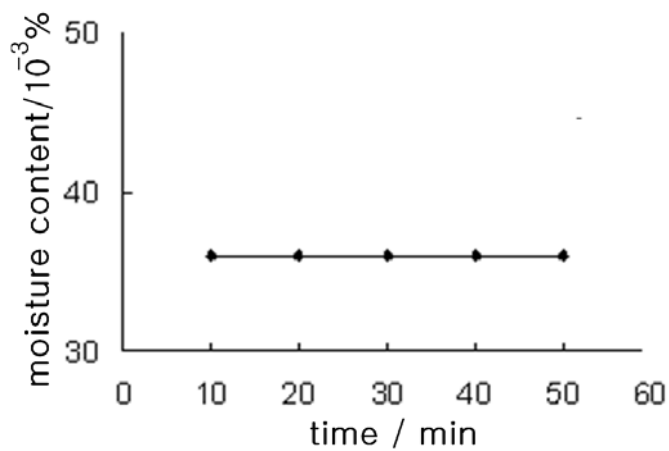

Figure 4. Change of the moisture content of the sample according to the measuring time

**Limit of measurement by the method of acetylene-production.** The change of peak area of acetylene according to the moisture content was measured. As the moisture content of the sample became increased, peak area of acetylene was increased linearly (Figure 5).

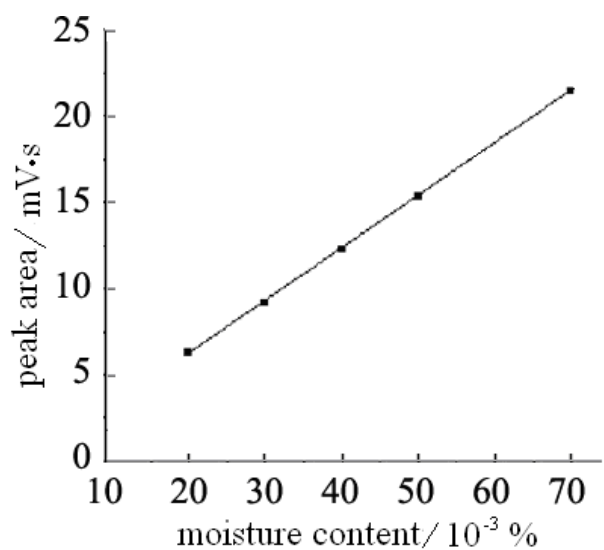

Figure 5. The change of peak area of acetylene according to the moisture content

Minimum limit of measurement by this method was $2 \cdot 10^{-3}\%$ because the peak area of noise by detector was 30 mV·s,

**Determination of moisture content of the objective sample.** To verify this method of standard addition, the moisture content of one of DMA samples was determined (Table 1), which showed that trace moisture in the DMA can be determined by this method with RSD of 2.272%.

Table 1.  Determination of trace moisture in DMA sample

| No. | primitive sample | | water (1.0 μL) added sample | | moisture content / $10^{-3}\%$ |
|---|---|---|---|---|---|
| | mass / g | peak area / mV·s | mass / g | peak area / mV·s | |
| 1 | 0.579 | 0.70 | 0.0031 | 11.13 | 36.0 |
| 2 | 0.576 | 0.70 | 0.0028 | 11.10 | 35.2 |
| 3 | 0.579 | 0.71 | 0.0029 | 11.10 | 34.3 |
| 4 | 0.581 | 0.73 | 0.0032 | 11.86 | 36.1 |
| 5 | 0.583 | 0.79 | 0.0027 | 11.02 | 36.2 |
| Mean ± S.D. ($\cdot 10^{-3}\%$) | | | | | 35.56±0.808 |
| R.S.D. (%) | | | | | 2.272 |

■ **CONCLUSIONS**

Here presented was a method for determination of trace moisture in DMA by acetylene production-gas chromatography. In the condition of 30℃ and ultrasonic dispersion, moisture in DMAe was reacted with calcium carbide to produce acetylene, which was determined by gas chromatography. By this method, the moisture content in DMA can be determined with RSD of 2.272% and limit of measurement of $2 \cdot 10^{-3}\%$.

■ **AUTHOR INFORMATION**

Corresponding Author

* ryongnam9@yahoo.com

■ **REFERENCES**